\documentstyle[12pt]{article}

\begin{document}
 
\vglue 3cm


\begin{center}
\vglue 0.5cm
{\Large\bf Topological Symmetry Breaking \\on Einstein Manifolds} 
\vglue 1cm
{\large A.Sako} 
\vglue 0.5cm
{\it Department of Physics,Hokkaido University,Sapporo 060 ,Japan}
\baselineskip=12pt

\vglue 1cm
{\bf ABSTRACT}
\end{center}
{\rightskip=3pc
 \leftskip=3pc}

\noindent
It is known that if gauge conditions have Gribov zero modes, then 
topological symmetry is broken. 
In this paper we apply it to topological gravity in dimension $n \geq 3$.   
Our choice of the gauge condition for conformal invariance is $R+{\alpha}=0$
, where $R$ is the Ricci scalar curvature.
We find when $\alpha \neq 0$, topological symmetry is not broken, 
but when $\alpha =0$ and solutions of the Einstein equations exist then 
topological symmetry is broken. 
This conditions connect to the Yamabe conjecture. Namely negative constant scalar
 curvature exist on manifolds of any topology, but existence of 
nonnegative constant scalar curvature is restricted by topology. 
This fact is easily seen in this theory.   
Topological symmetry breaking means that BRS symmetry breaking 
in cohomological field theory. 
But it is found that another BRS symmetry can be defined and physical 
states are redefined. The divergence due to the Gribov zero modes 
is regularized, and the theory after topological symmetry breaking 
become semiclassical Einstein gravitational theory 
under a special definition of observables.    
\vspace{3cm}

\begin{flushleft}
\baselineskip=12pt
\hrule
\vspace{0.2cm}
{sako@particle.phys.hokudai.ac.jp}
\end{flushleft}

\vfill\eject
\setcounter{page}{1}
\pagestyle{plain}
\baselineskip=14pt


\section{Introduction}

Topological field theories have been studied in these years  \cite{w-1} \cite
{P.R}. 
Especially, Seiberg Witten theory is making great development 
in 4-dim topology, now \cite{sw-1}. 
But there are few reports in which topological symmetry realize in our world. 
Hence, the symmetry should be broken in order to have some connection 
with our world \cite{tb} \cite{zl}.
In ref.\cite{zl}, Zaho and Lee add a infinitesimal breaking term to a Lagrangian. 
They found, that if gauge conditions are well defined 
and there is no Gribov zero mode 
then topological symmetry is hold in the limit of zero breaking term, 
but when there is a zero mode,
 BRS symmetry, i.e. topological symmetry, is broken. 
But the theory has such problems that physical 
meaning is lost after BRS symmetry 
breaking and divergence appear from Gribov zero modes. \\
We construct a topological gravitational theory of dimension $n \geq 3$ 
that has 
breaking phase with the method of Zaho and Lee.  
Topological gravitational theory we treat is fixed by $R+\alpha=0$ 
for the conformal symmetry. 
If we change the gauge condition as $\alpha \rightarrow 0$, then the 
theory become 
ill-defined by Gribov zero modes on some manifolds. Gribov zero modes appear 
when Fadeev-Popov matrix has zero eigen values.
In our theory, this zero eigenvalue equations are Einstein equations.   
Hence, topological symmetry is broken on only Einstein manifolds.
Strictly speaking, solutions of the Einstein equations exist, then 
 topological symmetry
is broken and smaller symmetry,  
 diffeomorphism-invariance, are left in this theory.
Further, we solve above problems. 
We recover the physical meaning to define 2nd BRS operator that is constructed
by the remaining symmetry, diffeomorphism-invariance. 
And one method of regularization to avoid divergence from zero modes 
is introduced in a general case. 
By using this regularization for gravitational 
theory, we get semiclassical Einstein gravitational theory in some case.\\ 
   On the physical point of view, our purposes are to treat Einstein manifolds
 and to construct quantum gravity of which classical limit become 
Einstein gravity. 
When cosmological constant is zero, then scalar curvature $R$ is zero on 
Einstein manifolds. If a gauge condition is $R=0$, Gribov zero modes 
might appear on some manifolds 
and theory would be ill-defined in former theory \cite{m-1} \cite{m-2}. 
In our theory, we are able to make the theory well-defined and having  
broken phase of topological symmetry at $R=0$. And this broken phase is 
interpreted as semi-classical gravity in a sense.\\  
The symmetry breaking conditions are connected to the Yamabe conjecture. 
We can easily find topological restriction to scalar curvature changing by 
conformal mode.  \\

This paper is organized as follows. 
 We review Zaho-Lee symmetry breaking theory\cite{zl} in a general case, in
section 2.
The core of this theory is to use singularity of Gribov zero modes. 
To avoid the zero modes, they added an infinitesimal breaking term to a  
Lagrangian.
Even though the breaking term is infinitesimal, 
it influence physical amplitude. 
In  section 3, we realize it in topological gravity.
As same as some other Witten type topological field theories\cite{P.R}
, topological gravity \cite{w-2}\cite{m-1}\cite{m-2} is constructed by BRS 
formalism .
We fix the conformal symmetry by $R+{\alpha}=0$
, where $R$ is scalar curvature and ${\alpha}$ is cosmological constant. 
Then if the Einstein equations and $R={\alpha}=0$ are simultaneously satisfied,
 topological symmetry is broken.
These conditions connect to the Yamabe conjecture \cite{Ym}.
Topological symmetry breaking means BRS symmetry breaking i.e. physical
structure lose its meaning, then.
In section 4 , we define 2nd BRS transformation. 
So we redefine physical states, 
then physical states recover with 2nd BRS operator after topological symmetry 
breaking.
In section 5, one method of reguralization for the zero modes is given in 
the general case including our gravitational case.
In section 6, we discuss about mathematical meaning.
The topological symmetry breaking conditions give some topological information
 to scalar curvature, which connect with the Yamabe conjecture here.
This information is due to the fact that absence of Gribov zero modes become 
a sufficient condition for functional subspace   
 be a infinite dimensional manifold.
In last section, we mention some conclusions, difficulties and prospects.


\section{General formalism}
%
\newcommand{\La}{{\cal L}}
\newcommand{\Lc}{{\cal L}_{\mit{cl}}}
\newcommand{\Dp}{{\cal D} \phi_k} 
\newcommand{\be}{\begin{equation}}
\newcommand{\de}{{\cal D}{^i}{_j}{_\epsilon}}
\newcommand{\De}{{\cal D}{^i}{_j}}
\newcommand{\ee}{\end{equation}}
\newcommand{\pBRS}{{\phi_j}\frac{\delta}{\delta \phi_j}}

Zaho, Lee showed that topological symmetry can be broken by  
singularity of Gribov zero mode \cite{zl}.
We review it and extend further to a general case in this section.
Let $\phi_i(x)$ represent all fields which include unphysical fields
 like ghost fields.
A total lagrangian $\La$ is represented by a classical Lagrangian $\Lc$ 
 , BRS operator $S$ , and gauge fermions $\Psi_{g.f}$, as 
\be{\La}={\Lc}+S{\Psi_{g.f}} \ee
 $\Psi_{g.f}$ is constructed with antighosts ${\bar{c}}^i$ , N-L fields
 $b_i$, and gauge fixing functions $F_i$. 
Namely our gauge conditions are $F_i=0$. 
In general, it  is possible to write 
 $\Psi_{g.f}=i{\bar{c}}^{i}(\alpha b_i +F_i)$, 
where $\alpha$ is a gauge parameter.
BRS transformation for   ${\bar{c}}^i$ and $b_i$ is defined by 
$S{\bar{c}}^i=b_i  , Sb_i=0 $. 
Then we have 
\be \La = \Lc + ib^i(\alpha b_i +F_i)+i{\overline{c}}^i(S\pBRS)F_i . \ee

We chose Landau gauge, $\alpha =0$, for simplicity.
We demand the total Lagrangian $\La$ has some BRS symmetry .
And we assume functional integral major $\Dp$ and a physical observable $O$
is invariant under BRS transformation. 
If there are Gribov zero modes, naive gauge fixing is not correct, 
so we need some
regularization to avoid it.
All gauge conditions $F_i$ do not have to include Gribov zero modes for 
symmetry breaking, but 
for simple notation we put regularization terms $i\epsilon{ b^i}{f_i}$ 
for each $F_i$ into the Lagrangian , 
\be \La_{\epsilon}=\La+i\epsilon{ b^i}{f_i}
            =\Lc + ib^i(F_i+\epsilon f_i)+i{\overline{c}}^i(S\pBRS)F_i ,\ee
 and demand that 
$i\epsilon{ b^i}{f_i}$ is not invariant under BRS transformation.
The vacuum expectation value of any observable $O$ is defined by
\be \langle O \rangle _\epsilon \equiv {\lim_{\epsilon \rightarrow 0}}\int \Dp 
\; O e^{-\int {dx}^D \La_{\epsilon}} .  \ee
If there is no singularity, then $i\epsilon{ b^i}{f_i}$ never influence,
\be \langle O \rangle _\epsilon =\langle O \rangle 
=\int \Dp \; O e^{-\int {dx}^D \La}  .\ee
In the following, I omit the index $\epsilon$ of 
 $\langle O \rangle _\epsilon$ and $\lim_{\epsilon \rightarrow 0}$ for
 convention. 
Now we estimate the vacuum expectation value of BRS exact functional 
$SO$ as,
\be \langle SO \rangle =\int \Dp\; SO e^{-\int {dx}^D \La_{\epsilon}}
   = \int \Dp \; O S(i\epsilon{ b^i}{f_i}) e^{-\int {dx}^D \La_{\epsilon}}.\ee
Since $SO$ is BRS exact, we usually expect it vanishing in the limit as 
$\epsilon $ approaches zero.
But if there are Gribov zero modes, $\langle SO \rangle \neq 0$ 
is realized as follows.
After $b_i$ integration in(5), we get
\begin{eqnarray} 
 \langle SO \rangle&=&-\epsilon \int \Dp \;\; 
O e^{\Lc+i\overline{c^i}(S\pBRS F_i)}\nonumber \\
&& {(S\pBRS{f_k})}
(\frac{\delta }{\delta F_k+\epsilon f_k}
\prod_{k}{\delta ( F_k+\epsilon f_k)})\nonumber \\  
\lefteqn{=-\epsilon \int \Dp \;\; O e^{-\int dx^D \Lc }
\prod_{k}{(S\pBRS{f_k})}}\\
&& (\frac{\delta }{\delta F_l+\epsilon f_l}
\prod_{m}{\delta ( F_m+\epsilon f_m)})
(i)^n (S\pBRS F_l) .\nonumber
\end{eqnarray}
Where we assume that the observable $O$ does not contain $b_i$ fields.
The second equality in eq.(7) was gotten by $\overline{c^i}$ integration, 
and $\phi_i$ represent all fields which were not still integrated.
Here, it turn out that when $\frac{\delta F_i}{\delta {\phi}_j}=0$ 
and $F_i=0$ are simultaneously satisfied then $\langle SO \rangle \neq 0$ from 
 $\frac{\delta }{\delta F_l+\epsilon f_l}
\prod_{m}{\delta ( F_m+\epsilon f_m)}$. 
It is easily understood by that $\delta (x^2+\epsilon )$ has strong divergence
as the limit $\epsilon \rightarrow 0$ and derivative $\frac{\delta }{\delta F_l+\epsilon f_l}$ go up the power of divergence. This is an essence of symmetry
 breaking. To see this apparently, 
next we change some of $\phi _i$ to gauge functions $F_i+\epsilon f_i$, 
and carry out their integral by using $ \delta ( F_k+\epsilon f_k)$.
We find 
\begin{eqnarray} \lefteqn{\langle SO \rangle=i^{n}\epsilon \int \Dp
\prod_{k} \frac{\delta }{\delta F_k+\epsilon f_k}
\{\frac{1}{|Det \frac{\delta(F_i +\epsilon f_i)}{\delta \phi_j}|}}\nonumber \\
&&\prod_{l,m}(S\pBRS{f_l}) (S\pBRS F_m) O e^{-\int dx^D \Lc } \}
\mid_{F_i +\epsilon f_i=0}. \end{eqnarray}
If $  { b^i}{f_i} $ did not break BRS symmetry i.e.$ S\pBRS{f_i} =0$ 
or $\de$ defined by  $\de = \frac{\delta(F_i +\epsilon f_i)}{\delta \phi_j}$ 
had no zero mode,
then $\langle SO \rangle=0$ in the limit $\epsilon \rightarrow 0$ . 
But now $b_if^i$ breaks the BRS symmetry , and we assume that 
there are some $\De$ zero modes at $F_i =0$ .
Then $\epsilon$ and $\de$ cancel each other. We get some non-zero value
$\langle SO \rangle \neq 0$ . This means that BRS symmetry is broken.
 
For example,

\be F_i +\epsilon f_i \mid _{\phi_j =\phi_j^c+\Delta\phi_j^c}=0
 \;\;\;\; F_i \mid_{ \phi_j =\phi_j^c}=0 \;\;\;\; j=1\sim {n} \ee
\be \De \mid_{\phi_j^c}= \frac{\delta(F_i )}{\delta \phi_j}
 \mid_{\phi_j^c}=0 \;\;\;\;   j=1\sim {n} \ee
for only one $\phi^c$. Where $n$ is a number of conditions $F_j$ 
and index ``$i$'' is fixed. In this case, as we will see in 
section 3, 
$\prod_i {\phi_j}\frac{\delta F_i}{\delta \phi_j} \sim \Delta \phi^c 
\sim \epsilon ^{\frac{1}{2}} $,
and  
${\de}^{-1} \sim \Delta \phi^c \sim \epsilon^{-\frac{1}{2}}$ 
when $f_i(\phi^c) \neq 0$.
 Then the most divergent term of 
 $\frac{\delta }{\delta F_k+\epsilon f_k}
 ({Det\de}^{-1})$  is order $ \epsilon^{-1}$  .
After all we get the order of the $\langle SO \rangle$ as,

\be \langle SO \rangle \sim \frac{\epsilon}{{\Delta \phi^c}^2}=1 \ee
We have seen some BRS symmetry is broken by Gribov zero modes in a 
general case.
If we use this way for Witten type topological field theories 
\cite{w-2}\cite{P.R}, then topological symmetry breaking may be realized.
\\

There are some problems of this method.
First, after BRS symmetry was broken, physical states lost their 
meaning. To solve this, we prepare 2nd-BRS operator for topological gravity
in section 4.
Second problem is whether partition function $Z$ is finite or not. 
At first sight it will be divergent. 
But we will see that it isn't true.

\begin{eqnarray} Z&=& \int \Dp e^{-\int {dx}^D \La_{\epsilon}} \nonumber \\
&=& \int \Dp e^{-\int {dx}^D \Lc}
(\frac{\delta(\phi_j -\phi_k^c-\Delta\phi_k^c)}{|Det\de |}
\prod_{l}S\pBRS{F_l}) \end{eqnarray}
Since $\de \sim \prod_{l}S\pBRS{F_l} \sim {\epsilon}^{\frac{1}{2}}$ , they cancel each other. 
So, the partition function keep finite. 
Third, as we see in (9), amplitude of some observable is divergent. 
This fact demand the theory to be regularized. 
We will give one method of regularization in section 5. 
\\


\section{ The case of Topological gravity}


\newcommand{\w}{{\sl w }}
\newcommand{\eR}{\frac{\delta e(R+\alpha +\epsilon f)}{\delta e^a_{\mu}}}
\newcommand{\wR}{\frac{\delta e(R+\alpha +\epsilon f)}{\delta w^{ab}_{\mu}}}
\newcommand{\eeR}{\frac{{\delta}^2 e(R+\epsilon f)}
{\delta e^a_{\mu} \delta e^b_{\nu}}}
\newcommand{\ewR}{\frac{{\delta}^2 e(R+\epsilon f)}
{\delta e^a_{\mu} \delta w^{bc}_{\nu}}}
\newcommand{\wwR}{\frac{{\delta}^2 e(R+\epsilon f)}
{\delta w^{ab}_{\mu} \delta w^{cd}_{\nu}}}
\newcommand{\eer}{\frac{{\delta}^2 \;eR}{\delta e^a_{\mu} \delta e^b_{\nu}}}
\newcommand{\ewr}{\frac{{\delta}^2 \;eR}{\delta e^a_{\mu} \delta w^{bc}_{\nu}}}
\newcommand{\wwr}{\frac{{\delta}^2 \;eR}{\delta w^{ab}_{\mu} \delta w^{cd}_{\nu}}}

We show here the theory of the previous section will be realized
 in topological gravity \cite{w+l}\cite{w-2}.
We use Myers theory \cite{m-2} that treat spin connection 
and vierbein as independent fields. Without this property, this theory
is almost same as Myers-Periwal theory \cite{m-1}. 
In our theory, dimension of the manifold is not essential as far as dimension
$n \geq 3$,  
but 4-dim case is quoted often for a simple example.  
In these theories, there are BRS operator $S$ and non nilpotents BRS-like operator
$\hat{S}$ which is reduced local orthogonal-transformations 
and diffeomorphism from $S$.
The $S$ is defined as 
\begin{eqnarray} 
Se^a_{\mu}&=&-({\w^a}_{b} + {P^a}_{b})e^b_{\mu}+ L_c e^a_{\mu},
\nonumber \\
S{\w^a}_b&=&L_c {\w^a}_{b} -{\w^a}_{c}{\w^c}_{b}-{P^a}_{c}{\w^c}_{b}
-{\w^a}_{c}{P^c}_{b}
-(L_{\phi}e_b^{\mu})e^a_{\mu}-{Q^a}_{b} , \nonumber \\
S{{w_{\mu}}^a}_b&=&{{\Lambda_{\mu}}^a}_b +{\nabla}_{\mu} {P^a}_{b}
+c^{\nu}{{R_{\nu \mu}}^a}_b , \nonumber \\
S{{\Lambda_{\mu}}^a}_b&=&L_c{{\Lambda_{\mu}}^a}_b-{\nabla}_{\mu}{Q^a}_{b}
-{{\Lambda_{\mu}}^a}_c {P^c}_b -{P^a}_c{{\Lambda_{\mu}}^c}_b
-\phi^{\nu}{{R_{\nu \mu}}^a}_b , \nonumber \\
Sc^{\mu}&=&c^{\nu}{\partial}_{nu}c^{\mu}+\phi^{\mu} , \nonumber \\
S\phi^{\mu} &=& L_c \phi^{\mu} , \nonumber \\
S{P^a}_b&=& -{P^a}_c{P^c}_b +{Q^a}_b- c^{\mu}{{\Lambda_{\mu}}^a}_b
+\frac{1}{2}c^{\mu}c^{\nu}{{R_{\mu \nu}}^a}_b ,\nonumber \\
S{Q^a}_b &=& L_c{Q^a}_b+{Q^a}_c{P^c}_b-{P^a}_c{Q^c}_b+\phi^{\mu} {{\Lambda_{\mu}}^a}_b
, \nonumber \\
Sx&=&y+L_cx-\delta_{P}x , \nonumber \\
Sy&=& L_cy-\delta_{P}y-L_{\phi}x+\delta_{Q}x .
\end{eqnarray}
%
Where $L_c , L_{\phi}$ denotes the Lie derivative for a fermionic vector field $c^{\mu}$
and for a bosonic vector field $\phi^{\mu}$. 
Also, $\delta_{P}, \delta_{Q}$ denote
  local orthogonal transformations by $P$ and $Q$.
$\phi^{\mu}$ and ${Q^a}_b$ are second stage ghosts for ghosts
 $c^{\mu}$ and ${P^a}_b$ in the Batalin, Fradkin and Vilkovisky formalism \cite{BFV}. 
 And $x$ and $y$ stand for all antighosts and N-L fields. \\
Myers and Periwal induce $\hat{S}=S-L_c+\delta_{P}$ and then $\hat{S}$ cohomolgy
 represent physical states. After straight forward calculation, we get 
${\hat{S}}^2=L_{\phi}+\delta_{Q}$. Then, for any scalar functional $h$ 
up to a total derivative, 
\begin{eqnarray}
Sh=\hat{S}h.
\end{eqnarray}

We will show that this $S$ symmetry is broken by the way of section 2.
We fix the GL transformation up to diffeomorphism and local orthogonal 
transformations at the same
conditions as Myers and  Periwal \cite{m-1}\cite{m-2}, in 4-dim case.
\begin{eqnarray}
R+\alpha &=& 0\\
{W^{+}}_{abcd} &=& 0 \\
\nabla_{\mu} e^a_{\nu}-\nabla_{\nu} e^a_{\mu} &=& 0 
\end{eqnarray}
While we choose the constraints to fix the redundant diffeomorphism and orthogonal transformations
\begin{eqnarray}
\nabla^{a}t_{ab}-\frac{1}{2}\nabla_{b}t=0 \; ,\; {r^a}_b=0, 
\end{eqnarray}
where $t_{ab}=\frac{1}{2}( {\w_{ab}}+{\w_{ba}}), \, t=tr(t_{ab}) ,\,
r_{ab}=\frac{1}{2} ( {\w_{ab}}-{\w_{ba}})$. 
These conditions are all covariant. 
So diffeomorphism and local orthogonal transformations are still unfixed. 
In ref.\cite{m-2}, to fix these symmetries, they imposed harmonic condition 
$\partial_{\mu}(e \,e_a^{\mu}e^a_{\nu})=0$ and algebraic constraint 
$\tilde{e}^{\mu}_{[a}e_{b]\mu}=0$ where 
$\tilde{e}^{\mu}_{a}$ is some fixed back ground tetrad. \\
But now, we do not adopt these conditions and  
the reason will appear in the next section. 
So, we assume these symmetries were fixed by some appropriate conditions. \\

Let us adapt this topological gravity to the way of section 2. 
On the Landau gauge, we get delta functions after N-L fields integration. 
According to the previous section, some symmetry breaking term should be added. 
We take it $i\epsilon e\tau f(e^a_{\mu},{w_{\mu}}^{ab})$, 
where $f$ is some functional of $e^a_{\mu}$ and ${w_{\mu}}^{ab}$
that satisfy $Sf\neq0$ , and $\tau$ is an N-L field used to fix on condition (15). We are able to regard eq.(15) as a fixing condition for conformal mode. 
Additional gauge fixing Lagrangian for eq.(15) can be written with antighost $\rho$ and 
its N-L fields $\tau =\hat{S}\rho$ as, 
\begin{eqnarray}
\La_{\alpha}&=&Se\rho (R+\alpha) \nonumber \\
            &=&\hat{S}e\rho (R+\alpha) \nonumber \\
            &=&e\tau (R+\alpha) -\rho \hat{S}(e(R+\alpha))
\end{eqnarray}
 
The formula (14) was used for second equality of (19), and total divergence
was ignored.
Due to the additional symmetry breaking term, the delta function changes from 
$\delta(e(R+\alpha))$ to $ \delta(e(R+\alpha +\epsilon f))$ . 
So, in this case, symmetry breaking is only connected to the 
condition $R+\alpha =0$ . 
We abbreviate another conditions for redundant diffeomorphism 
and local orthogonal symmetry, like eq.(18), 
and GL symmetry without conformal mode
, like eq.(16) and (17) to ``$ (\; GL \;)$''.  
The total Lagrangian is written as below by using appropriate gauge fermions
 $x_1(diffeo.)+x_2(ortho.)$ for fixing the diffeomorphism and local
orthogonal transformations, 
 
\begin{eqnarray}
\La_\epsilon &=& \Lc + \La_{\alpha} +\epsilon e\tau f
             +{\hat{S}}x_0(\;GL\;)+Sx_1(diffeo.)+Sx_2(ortho.) \nonumber \\
   &=& \Lc +e\tau (R+\alpha+\epsilon f) +\rho \hat{S}(e(R+\alpha)) \nonumber \\
             & & +{\hat{S}}x_0(\;GL\;)+Sx_1(diffeo.)+Sx_2(ortho.)
\end{eqnarray}
where $x_0,x_1$ and $x_2$ are antighost fields and their tensor property 
is determined by gauge functions $(\;GL\; ),(diffeo.)$ and $(ortho.)$.
We get delta functions after N-L fields $y$ integration 
\be
\delta(e(R+\alpha+\epsilon f) \prod \delta (\;GL\; )\delta (diffeo.)\delta (orth.)
\ee
For simplicity, all delta functions from 
$ {\hat{S}}x_0(GL)+Sx_1(diffeo.)+Sx_2(ortho.)$
 are denoted by $\prod \delta (\;GL\; )\delta (diffeo.)\delta (orth.)$, here.
The number of these delta functions is the same number of components of 
$e^a_{\mu}$ and $ w^{ab}_{\mu}$, because topological symmetry permit to transform 
each components arbitrary.
So, if $(\;GL\; )$,$(diffeo.)$ and $(orth.) $ contain only $e^a_{\mu} \,$ and $ \,  w^{ab}_{\mu}$
, we can rewrite (21) as 
\begin{eqnarray}
\lefteqn{ \delta(e(R+\alpha+\epsilon f) 
\prod \delta (\;GL\; )\delta (diffeo.)\delta (orth.)}\\   
& &= {\cal J}^{-1}\prod_{a,b,c,\mu ,\nu } \delta (e^a_{\mu}-e^{a(c)}_{\mu}-\Delta e^{a(c)}_{\mu})
 \delta (w^{bc}_{\nu}-w^{bc(c)}_{\nu}-\Delta w^{bc(c)}_{\nu}). \nonumber
\end{eqnarray}
Where ${\cal J}$ is Jacobian, 
\begin{eqnarray}
{\cal J}={ \left| \begin{array}{cc}
{\eR}&{\frac{\delta (GL,diffeo.,orth.)}{\delta e^a_{\mu}}}\\
{\wR}&{\frac{\delta (GL,diffeo.,orth.)}{\delta w^{ab}_{\mu}}}\\
\end{array}\right|}, \nonumber 
\end{eqnarray}
$e^{a(c)}_{\mu}$ , and $w^{bc(c)}_{\nu}$ are solution i.e. 
\begin{eqnarray}
e(R+\alpha) | _{e^a_{\mu}=e^{a(c)}_{\mu},w^{bc}_{\nu}=w^{bc(c)}_{\nu}}=0 
\nonumber \\
(GL,diffeo.,ortho.)| _{e^a_{\mu}=e^{a(c)}_{\mu},w^{bc}_{\nu}=w^{bc(c)}_{\nu}}=0 ,
\end{eqnarray}
and $\Delta e^{a(c)}_{\mu}$ $\Delta w^{bc(c)}_{\nu}$ are variation from 
inducing $\epsilon f$. Note that they are depend on $\epsilon$ . 
 As we saw in section 2, if the Jacobian ${\cal J} $ has zero modes, then BRS symmetry, i.e. topological 
symmetry, is broken.\\
 
Let us analyze these broken conditions further. 
 We analyze the situation that each component of the first column 
of the Jacobian matrix vanishes, 
\begin{eqnarray}
\eR\vert _{e^{a (c)}_{\mu}}&=&e\,e_a^{\mu}(R+\alpha )+e{R^{\mu}}_{a} |_{e^{a(c)}_{\mu}} =0.\\ 
\wR &=&-6e \, e^{\mu}_{[a} e^{\nu}_l e^{\lambda}_{b]} (D_{\nu} e^l_{\lambda})=0
\end{eqnarray}

Eq.(24) is Einstein equation with cosmological constant $\alpha $. 
From eq.(23), eq.(24) become
\be
{R^a}_{\mu} |_{e^{a(c)}_{\mu}} =0. 
\ee

So we conclude that only if $R=\alpha =0$ and eq.(25) are satisfied, 
the topological symmetry is broken. 
In Calab-Yau manifolds in 6-dim  the conditions are satisfied, for example. 
But in many manifolds, they are not satisfied. 
This condition connects to Yamabe conjecture, and it will be discussed
 in section 6. 

The equations (25) mean torsion free conditions and these are not contradictory 
to gauge conditions if we adopt (17).  
\\
When these conditions are satisfied, Jacobian is of  
order $\epsilon ^{\frac {1}{2}}$. 
Indeed 
$e(R+\epsilon f)$ can be expanded around $e^{a(c)}_{\mu}$ as follows, 
\begin{eqnarray}
0&=&e(R+\epsilon f)|_{e^{(c)}+\Delta e^{(c)}\; , \; w^{(c)}+\Delta w^{(c)}} \\
&=&\epsilon e\,f +
\epsilon \frac{\delta e f}{\delta e^a_{\mu}}\Delta e^{a(c)}_{\mu} + 
\epsilon \frac{\delta e f}{\delta w^{ab}_{\mu}}\Delta w^{ab(c)}_{\mu} +
\eeR \Delta e^{a(c)}_{\mu}\Delta e^{b(c)}_{\nu} \nonumber \\
&+& \ewR \Delta e^{a(c)}_{\mu}\Delta w^{bc(c)}_{\nu}+
\wwR \Delta w^{ab(c)}_{\mu}\Delta w^{cd(c)}_{\nu}+
O(\Delta e^{3}). \nonumber 
\end{eqnarray}
To leading order in $\epsilon $ we have ,
\begin{eqnarray} 
\epsilon &=&{\eer}(ef)^{-1}\Delta e^{a(c)}_{\mu}\Delta e^{b(c)}_{\nu} 
+ {\ewr}(ef)^{-1} \Delta e^{a(c)}_{\mu}\Delta w^{bc(c)}_{\nu} \nonumber \\
& &+{\wwr}(ef)^{-1}\Delta w^{ab(c)}_{\mu}\Delta w^{cd(c)}_{\nu} 
\end{eqnarray}
in $f|_{e^{a(c)}_{\mu}} \neq 0$ case. 
Hence, $\Delta e^{a(c)}_{\mu}$ and $\Delta w^{ab}_{\mu}$ 
should be order $\epsilon ^{\frac{1}{2}}$.\\ 
We can estimate (24) as,  
\begin{eqnarray}
\lefteqn{\frac{\delta e(R+\epsilon f)}{\delta e^a_{\mu}}
|_{e^{(c)}+\Delta e^{(c)}, \; w^{(c)}+\Delta w^{(c)}}}\\
& &= \epsilon \frac{\delta e f}{\delta e^a_{\mu}} + 
\eeR \Delta e^{b(c)}_{\nu} 
 + \ewR \Delta w^{bc(c)}_{\nu}+
O(\Delta e^{2}) \nonumber
\end{eqnarray}
Now, we get $\frac{\delta e(R+\epsilon f)}{\delta e^a_{\mu}}
\sim  \epsilon ^{\frac{1}{2}}$ ,and similarly 
$\frac{\delta e(R+\epsilon f)}{\delta w^{ab}_{\mu}} \sim  \epsilon ^{\frac{1}{2}}$. 
From the estimation described above, it is concluded that 
Jacobian ${\cal J} $ is order $\epsilon ^{\frac{1}{2}}$ when $\alpha$ is 0
and a solution of the eq.(26), $R^a_{\mu}=0$,
 exist without contradiction to gauge
conditions, 
then topological symmetry is broken. 
Note that the singularity from the Gribov zero modes, Jacobian matrix zero 
eigen values, is contribution from 
only $\delta(e(R+\epsilon f))$. Even if we did not use Jacobian 
${\cal J}$ to estimate 
the singularity, we could find that the symmetry breaking occur 
by only this delta function. 
\\

 In this section, we have studied only about vacuum condensation in topological
gravity by the method of section 2.
We found that the topological symmetry breaking was appeared 
in the process of conformal changing $R\rightarrow 0$ 
if Ricci flat (26) is realized on the background manifold.  
Of course, for this theory being well defined as physical theory, 
some other BRS symmetry should be present. 
This is a subject of the next section.


\section{ Two-BRS formalism }

In section 2, we saw topological symmetry was broken at $R=0$. 
However this means that BRS quantization is ill-defined. 
To clear this problem, we introduce another BRS transformation. 
$L_c$ operates as diffeomorphism to all fields except anti ghosts.
Anti ghost fields were transformed as scalar fields, regardless of 
those tensor property. 
As a result of this, 
in our Lagrangian $\La=\Lc+\hat{S} \Psi+S{x_1}(diffeo.)+S{x_2}(ortho.)$
where $\hat{S} \Psi=\La_{\alpha}+\hat{S}x_0(GL)$, 
$\Lc$ and $\hat{S}$ exact gauge fixing term $ \hat{S} \Psi_{g.f.}$
 are $L_c$ invariant up to total divergence, 
because these terms are scalar, ref.\cite{Fujikawa}\cite{m-1}\cite{m-3}.
 So, if we can chose $S{x_1}(diffeo.)+S{x_2}(ortho.)$ to be 
invariant under $L_c$  and redefine $L_c$ to be nilpotents, then we adopt
$L_c$ as new BRS operator and physical states can be redefined with it. 

 Now one defines appropriate anti ghosts $x$ and Lagrange multipliers $y$
, where  
\be
 S{x_i}={y_i} \;\; S{y_i}=0 \;\;\; i=1,2.   
\ee
The tensor properties of these ${x_i}$ and ${y_i}$ are
 determined by $(diffeo.)$ and $(ortho.)$.
One can require $(diffeo.)$ is scalar under local orthogonal transformations
 but it has no general covariant property,
and $(ortho.)$ is scalar under general coordinate transformations.  
Under this choice, $S{x_1}(diffeo.)+S{x_2}(ortho.)$ is 
\begin{eqnarray}
& &S{x_1}(diffeo.)+S{x_2}(ortho.) \nonumber \\
& &= (L_c +\hat{S}){x_1}(diffeo.)
+(\delta _P +\hat{S}){x_2}(ortho.) \nonumber \\
& &= {y_1}(diffeo.)+ (-)^{|x_1|}{x_1}(L_c +\hat{S})(diffeo.) \nonumber \\
& &{\;\; }+ {y_2}(ortho.)+(-)^{|x_2|}{x_2}(\delta _P +\hat{S})(ortho.)
\end{eqnarray}

If we want to regard $L_c $ ( or $\delta _P$) as a new BRS operator,
it is seen in (31) that gauge fermions $(diffeo.)$ ( or $(ortho.)$)
should be $\hat{S}$ cohomology, and $ L_c {x_i}={y_i} $ 
( or $ \delta_P {x_i}={y_i}$). 
But $\hat{S}$ cohomological gauge conditions are not known at least to us. 
Only we know $\hat{S} \phi_{\mu} =0$ where $ \phi_{\mu}$ was induced as 
ghost for ghost $c_{\mu}$ in (13), ref.\cite{m-1}. 
So we adopt some functional of only $\phi_{\mu}$ for gauge conditions 
$G_{\widetilde{\mu \nu}}(\phi_{\mu})=0$, where tileder means 
$G_{\widetilde{\mu \nu}}$ is not general covariant. 
Note that $\hat{S}G_{\widetilde{\mu \nu}}=\frac{\delta}{\delta \phi _{\mu}}
G_{\widetilde{\mu \nu}}\hat{S} \phi _{\mu}= 0$, 
and $\delta_P G_{\widetilde{\mu \nu}}=0$ since 
$G_{\widetilde{\mu \nu}}$  has no local coordinate index. 
To fix the diffeomorphism, we add to Laglangian 
with anti ghost ${\bar{c}}^{\mu \nu }$ 
and Lagrange multiplier ${b}^{\mu \nu }=S {\bar{c}}^{\mu \nu }$
\be
Se\, {\bar{c}}^{\mu \nu } G_{\widetilde{\mu \nu}}
=et{\bar{c}}^{\mu \nu } G_{\widetilde{\mu \nu}} 
+(L_c e){\bar{c}}^{\mu \nu } G_{\widetilde{\mu \nu}} 
+et{b}^{\mu \nu } G_{\widetilde{\mu \nu}} 
+et{\bar{c}}^{\mu \nu } L_cG_{\widetilde{\mu \nu}} 
\ee
Next step, we fix the local orthogonal symmetry. 
Myers-Periwal fixed it at $\tilde{e}^{\mu}_{[a}e_{b]\mu}=0$, where
$\tilde{e}^{\mu}_{a}$ is some back ground tetrad. 
This condition is not suitable for our purpose. 
Because, under $L_c$, $\tilde{e}^{\mu}_{a}$ do not transform , 
so  $\tilde{e}^{\mu}_{[a}e_{b]\mu}$ is not invariant.
i.e. $L_c e{\bar{P}}^{ab} \tilde{e}^{\mu}_{[a}e_{b]\mu}\neq 0$, 
where ${\bar{P}}^{ab}$is anti ghost.
For this reason, another condition that include no back ground field, 
should be induced, here. 
For example, we fix it at ${\nabla}_{\mu}{{w^{\mu}}_{ab}}=0$ \cite{no}.
The gauge fixing terms, 
\be
Se{\bar{P}}^{ab}{\nabla}_{\mu}{{w^{\mu}}_{ab}}
=et{\bar{P}}^{ab}{\nabla}_{\mu}{{w^{\mu}}_{ab}}
+e{q}^{ab}{\nabla}_{\mu}{{w^{\mu}}_{ab}}
-e{\bar{P}}^{ab}(\hat{S} +\delta _p){\nabla}_{\mu}{{w^{\mu}}_{ab}}, 
\ee
where ${q}^{ab}$ is a Lagrange multiplier, 
${q}^{ab}=S{\bar{P}}^{ab}$, is added to Lagrangian.  
\\
Everything is ready, for introducing new BRS symmetry. 
Let us define a new fermionic Lie derivative $L_c'$ for some BRS operator.
\newtheorem{definition}{definition}
\begin{definition}[new BRS operator $L_c'$] The new BRS operator is defined
as follows,  
\be 
L_c'=L_c
\ee
\centering{ for all fields $e^a_{\mu},{\w^a}_b,w^{ab}_{\mu},\dots$ except $\bar{c}_{\mu \nu} , \; b_{\mu \nu}$}
\begin{eqnarray}
L_c'\bar{c}_{\mu \nu}&=& t \bar{c}_{\mu \nu} +b_{\mu \nu} \\
L_c'b_{\mu \nu}&=& tb_{\mu \nu} -(L_c t)\bar{c}_{\mu \nu}
\end{eqnarray}
\centering{  for $\bar{c}_{\mu \nu} , \; b_{\mu \nu}$. }
\end{definition}

Here $L_c'$ is nilpotents. 
From (34), $\Lc$ , $ \hat{S}\Psi$ and $Se{\bar{P}}^{ab}{\nabla}_{\mu}{{w^{\mu}}_{ab}}$
are transformed as scalar by $L_c'$, 
that is  
\be
\int d^D x L_c'(\Lc + \hat{S}\Psi 
+ Se{\bar{P}}^{ab}{\nabla}_{\mu}{{w^{\mu}}_{ab}})=0.
\ee 
And  by using (36),
\be
 Se\, {\bar{c}}^{\mu \nu } G_{\widetilde{\mu \nu}}
=L_c'e\, {\bar{c}}^{\mu \nu } G_{\widetilde{\mu \nu}},
\ee
we get 
\be
L_c'\; Se\, {\bar{c}}^{\mu \nu } G_{\widetilde{\mu \nu}}=
(L_c')^2 e\, {\bar{c}}^{\mu \nu } G_{\widetilde{\mu \nu}}=0, 
\ee
from nilpotency of $L_c'$.
Our total action is rewritten with $L_c'$, as 
\be
\int d^D x \; (\Lc + \hat{S}\Psi 
+ Se{\bar{P}}^{ab}{\nabla}_{\mu}{{w^{\mu}}_{ab}})+
L_c'e\, {\bar{c}}^{\mu \nu } G_{\widetilde{\mu \nu}}
\ee
and it is invariant under nilpotents operator $L_c'$ as we saw in (37) and (39).
Now it is possible to regard $L_c'$ as a new BRS operator, and
 $  \Lc + \hat{S}\Psi + Se{\bar{P}}^{ab}{\nabla}_{\mu}{{w^{\mu}}_{ab}}$
is a new classical Lagrangian.
In this form, ${t^a}_b , \phi _{\mu} $ and others except 
${\bar{c}}^{\mu \nu }$ and ${b}^{\mu \nu }$ become physical fields
in addition to $e^{a}_{\mu} $ and $ w^{ab}_{\mu}$, 
as a result of changing the 
physical states conditions from $S|phys>=0$ to $L_c'|phys>=0$.\\

Note that using $\phi _{\mu}$ to fix diffeomorphism disturbs rewriting 
$\delta$ functions with Jacobian like the previous section.
Because the number of $\delta$ functions of $e^{a}_{\mu} , w^{ab}_{\mu}$
 is less than one of independent components of fields. 
But singularity of the $\delta (eR)$ is not changed. 
It is apparent that topological symmetry is broken.\\

The symmetry breaking condition is in fact the Einstein equation  $R_{ab}=0$. 
The  solution of a classical Einstein equation exists 
 and it contributes to the path integral
, hence the theory of broken topological symmetry 
is realized in the above. 
In other words, non-topological phase is defined around classical
gravity.


\section{Regularization}

In the previous section, we got the new BRS operator to define physical states again. 
It means that the vacuum expectation value of some BRS exact operator is zero. 
But our theory is still singular due to $\delta (e(R+\epsilon f))$ in (21). 
So we have to remove this divergence. 
In this section, we make the prescription to regularize this divergence. \\

First, we discuss reguralization in general formalism. 
Here, we use same symbols as ones of section 2. 
There is strong divergence in $\delta (F_i+\epsilon f_i)$ in eq.(7)
as $\epsilon \rightarrow 0$ when $\De (\phi^c)=0$. 
This delta function appeared as a result of $b_i$ integration,
 where $b_i$ is a Lagrange multiplier of the gauge function $F_i$. So it is evident 
that more stronger divergence will appear if an observable contains $b_i$ 
fields. 
To get finite vacuum expectation value, we redefine $b_i$ fields as 
\be 
{b'}_i = {\epsilon}^{\frac{1}{2}}b_i
\ee
 in $f_i(\phi^c)\neq 0$ case when 
\be
\frac{\delta F_i}{\delta {\phi}_j} {\mid}_{\phi^c}=\De (\phi^c)=0 \;\;\;\;
\frac{\delta F_k}{\delta {\phi}_j} {\mid}_{\phi^c} \neq 0 \;\;\;: k \neq i \; ,\;
j=1 \sim n 
\ee 
are satisfied for some $\phi^c$. 
A fixed index ``$i$'' means that functional derivative of $F_i$ on some 
 $\phi^c$ vanish like eq.(42) in the following.
$\phi^c$, i.e. solutions of $F_j=0$, do not always satisfy eq.(42), 
then we put $\phi^z$ and $\bar{\phi^z}$ as 
\be
 \phi^z \in \{ \phi^c \mid \De (\phi^c)=0 \} \ \ \ 
\bar{\phi^z }\in \{ \phi^c \mid \De (\phi^c)\neq 0 \}
\ee
In other words, $\phi^z$ is a kernel of $\De$.
By using this ${b'}_i$, we rewrite the gauge fixing term in the Lagrangian (3) as 
\be 
b^i(F_i+\epsilon f_i) \rightarrow \epsilon^{-\frac{1}{2}}
{b'}^i( F_i+\epsilon f_i).
\ee
Then the delta function is changed as     
\be 
\delta (F_i+\epsilon f_i)\rightarrow \delta (\epsilon^{-\frac{1}{2}}(
F_i+\epsilon f_i))= \epsilon^{\frac{1}{2}} \delta (F_i+\epsilon f_i)
\ee
Due to eq.(45), delta functions are order 1 for $\phi^z$, 
but for $\phi^c =\bar{ \phi^z}$ they are order $\epsilon^{
\frac{1}{2}}$. 
In the count of the term $\prod_m S\phi_j \frac{\delta}{\delta \phi_j} F_m$ 
as similar in eq.(7) all amplitude of observables is order 
  $\epsilon^{\frac{1}{2}}$ whether there are zero modes or not, as it is. 
So, also $\bar{c^i}$ fields have to be redefined as 
\be
\bar{{c'}_i}= \epsilon^{\frac{1}{2}} \bar{c_i} 
\ee 
and the Fadeev Popov terms in the Lagrangian change as 
\be
 \bar{c^i} S\phi_j \frac{\delta}{\delta \phi_j} F_i=\bar{{c'}^i}
(\epsilon^{-\frac{1}{2}}  S\phi_j \frac{\delta}{\delta \phi_j} F_i).
\ee
As a result of these redefinition, order of $\epsilon^{\frac{1}{2}} \delta (F_i+\epsilon f_i)$ and $\prod_m S\phi_j \frac{\delta}{\delta \phi_j} F_m$ are 
found as 
\begin{eqnarray}
\epsilon^{\frac{1}{2}} \delta (F_i+\epsilon f_i) \sim \left\{
\begin{array}{c}
1 \ \ \colon \ \ {\rm for}\ \ \phi^z \\
\epsilon^{\frac{1}{2}} \ \ \colon \ \ {\rm for}\ \ \bar{\phi^z} 
\end{array}
\right. \\
\epsilon^{-\frac{1}{2}} S\phi_j \frac{\delta}{\delta \phi_j} F_i \sim \left\{
\begin{array}{c}
1 \ \ \colon \ \ {\rm for}\ \ \phi^z \\
\epsilon^{-\frac{1}{2}} \ \ \colon \ \ {\rm for}\ \ \bar{\phi^z}. 
\end{array}
\right.
\end{eqnarray}
Before these redefinitions, if an observable contains $b_i$ fields, then 
singularity is stronger. 
Indeed from the same reason, in section 2, $\langle SO \rangle$ 
was non zero, and yet partition function was finite. 
It is easy to understand by a following formulation of a delta function  
\be 
\lim_{\epsilon \rightarrow 0} \int db \; b e^{ib(x^2-\epsilon^{2})}
=\lim_{\epsilon \rightarrow 0} \frac{1}{2x}\frac{\partial}{\partial x}
\left( \frac{\delta (x-\epsilon) +\delta (x+\epsilon)}{|2x|} \right). 
\ee
Where existence of $b$ induced a derivative and power of divergence went up. 
But now, because of redefinition (41) and (46), 
when an observable contains ${b'}_i$  fields, i.e. $O={b'}_iO^i$ , 
then the vacuum expectation value of $O$ is 
\begin{eqnarray}
\langle  {b'}_iO^i \rangle = \int \Dp \; O^i \epsilon^{\frac{1}{2}} 
\left( \frac{\partial \ \ \delta (F^i+\epsilon f^i)}{\partial (F^i+\epsilon f^i)}
\right) 
 \prod_{k \neq i} \delta(F_k) \prod_m   S\phi_j \frac{\delta}{\delta \phi_j} F_m \; e^{\int dx^D \Lc }. 
\end{eqnarray}
Where the power of divergence is unchanged for $\phi^c=\bar{\phi^z}$ 
and the power of 
$\epsilon$ go up for non zero mode $\phi^c \neq \phi^z$. 
So that, contribution for amplitude is from only Gribov zero modes 
$\phi^z$, and other contribution from $\bar{\phi^z}$ vanish. 
After all, the amplitude (51) is sum over $\phi^z$ and it is order one because 
\be
 \frac{\partial \ \ \delta (F^i+\epsilon f^i)}{\partial (F^i+\epsilon f^i)} 
\sim \epsilon^{-1} + \;\; less\ \ divergence. 
\ee 
Hence, we have done the regularization for all observable in general case.
Especially there is remarkable property that if an observable contains Lagrange
multipliers of $F_i$ then contribution from $\phi_i$ path integration to 
the amplitude of this observable is only from $\phi^z$.   
Next we try this regularization in topological gravity case. \\

We carry out this regularization in the gravitational theory
 in the same way as the general case. 
The only things we have to do is to redefine $\rho$ and $\tau$ as follows, 
\be 
{\rho}'=\epsilon^{\frac{1}{2}} \rho \ \ \ \, \ \ \ \ {\tau }'=\epsilon^{\frac{1}{2}} \tau
\ee
Then all amplitude is regularized. 
Note that, if an observable contains ${\tau}'$, its vacuum expectation value 
is the sum of solution of $R^a_{\mu}=0$. 
This fact is very interesting. 
In the theory of section 4, we define observables as $L_c'$ closed. 
If we change this definition to  $L_c'$ closed and containing ${\tau}'$
fields, 
\begin{eqnarray}
Z&=& \int \Dp {\tau}'e^{\int dx^D \La_{\epsilon}} \ \ \colon \ \ {\rm 
partition function}\\
 L_c'O&=&0 \ \ {\rm and} \ \  O={\tau}'O' \ \ \colon \ \ {\rm definition \; of 
\; observable .} 
\end{eqnarray}
Then the theory is semiclassical, i.e.path integral contribution 
for vacuum expectation 
value is from only solution of the Einstein equation. 
 Note that our theory have many constraints for fixing topological 
symmetry like eq.(16) and (17). 
So, after symmetry breaking, these constraints are left as equations of motion. 
 In this meaning, sence of semiclassical gravity is different from usual case.


\section{ Mathematical interpretation }

As we saw in section 3, our theory has broken phase on  
the condition $R_{ab}=0$. Let us clarify  mathematical meanings of this. \\
For simplicity, we omit the symmetry breaking term 
 $\epsilon e\tau f$ in this section.
As is mentioned  in section 3, the Yamabe conjecture is concerned with our theory. 
We are going to see this fact, as follows.
In topological gravity, it is trivial that any physical amplitude is invariant
under changing $\alpha$ because gauge condition does not affect 
physical amplitude.
 Indeed, a  derivative of the partition function with respect $\alpha$
is given by,  
\be
\frac{\partial}{\partial \alpha} Z =\int \Dp \, (\hat{S} {ie \rho})
e^{\int dx^D \La}=0
\ee
To get the second equality, we use that  
$\hat{S}$ exact  vacuum expectation value vanishes as same as $S$ exact one, ref.
\cite{m-1} \cite{m-3}. 
This means that variation of scalar curvature is not restricted by topology.
Strictly speaking, our topological gravity may not classify the 
topology of manifolds perfectly, 
so we can only say that scalar curvature can be varied without changing class
 which is classified by our topological theory, as far as the theory is well defined.
But our theory is broken at $R_{ab}=R=0$ as we saw in section 3.
From a mathematical viewpoint, gauge conditions restrict back ground manifolds
to submanifolds.
The inverse mapping theorem demand that Jacobian ${\cal J} $ is nonzero for the neighborhood
 of the gauge conditions to be an infinite dimensional manifold. 
i.e. if the Jacobian is nonzero, then functional $R+\alpha ,(GL,diffeo.,ortho.)$
and their inverse map are isomorphism, and the inverse map
 $ \lbrace e^a_{\mu} , w^{ab}_{\mu} | R+\alpha=0 ,
(GL,diffeo.,ortho.)=0 \rbrace $
become manifold. 
If $R_{ab}=R=0$, then the Jacobian is zero and the submanifold is ill-defined.
 It is known that we can regard a topological field theory as an extended 
Morse theory \cite{laba}. In this point of view 
 if submanifolds are ill-defined then the Morse theory
 is also ill-defined and topological symmetry is breakdown.\\ 

This situation is similar in the Donaldson theory \cite{Donald} \cite{w-1}.
Homologies of moduli ${\cal M}=\lbrace \;[{\cal A} ] \; |\; F^+_{\cal A} =0 \rbrace$
can be defined, if and only if ${\cal M}$ is finite dimensional 
manifolds, where $[ {\cal A} ]=\lbrace {\cal A}/ {\cal G} \rbrace $ .
Donaldson invariants are Homology $H_*([{\cal A}];{\bf Z})$. 
So, they can be defined in the neighborhood of $F_{\cal A}^+ =0$ 
when the Jacobian $det(d^+_{{\cal A}}) \neq 0$, 
where $d^+_{{\cal A}}=d_{{\cal A}}+*d^+_{{\cal A}}$.
In other words, $d^+_{{\cal A}}$ should be surjection. \\
 In our theory, the condition that the Jacobian become zero is $R_{ab}=0$, 
and if it's satisfied topological symmetry is lost. \\

Yamabe conjectured constant scalar curvature $R$ exist on any  
compact Rieman manifolds with arbitrary topology of dimension 
$n \geq 3$ \cite{Ym}. 
But it has been corrected by Aubin \cite{Aubin} , Schoen \cite{Sch} and so on.
Especially, Kazdan and Warner \cite{K-W} gave the following theorem that
\newtheorem{theorem}{Theorem}
\begin{theorem}[Kazdan Warner theorem]
Compact manifolds $M$ of dimension $n \geq 3$ can be divided into three classes,\\
(A) Any $(C^{\infty})$function on $M$ is the scalar curvature of some  
$(C^{\infty})$metric. \\
(B) A function on $M$ is the scalar curvature of some metric if and only if
 it is either identically zero or strictly negative somewhere, further more,
 any metric with vanishing scalar curvature is Ricci-flat. \\
(C) A function on $M$ is a scalar curvature if and only if it is strictly
negative somewhere.
\end{theorem}

This theorem says that existence of negative scalar curvature do not demand 
any topological condition. 
And there is a barrier at $R=0$. This is consistent with our theory. 
We are able to classify the type (C) manifolds from the type (A) and
  (B) manifolds, in our theory.
Let us take $R=-\alpha < 0$ first, and makes $\alpha$ to zero. 
On the type (A)(B) manifolds, topological symmetry is broken ,
on the other hand, on the type (C) it's not broken. 
In other words, our theory may classify manifolds to type (C) and other type,
by calculating some vacuum expectation value of $\hat{S}$
exact terms on $R=0$. If it vanish, the back ground manifold is type(C), and if it's not zero, then 
the manifold is type (A)or(B).\\

Note that in Myers and Periwal \cite{m-1} observables are topological
on $\alpha \neq 0$.
On the other hand, they are topological-like but are non topological
in a strict sense in our theory. 
On type (C), they are independent of metric, but on type(A) or (B)
they are non topological.


\section{ Conclusion and discussion }

We have constructed a topological-like gravitational theory. Its feature is 
that the 
topological symmetry is broken when gauge condition chose $R=0$ and the Einstein 
equation $R_{ab}=0$ has a solution on the back ground manifolds. 

Now, the question flowing up naturally is how matter couples. 
For example, one change gauge condition to 
$R+\tilde{S}_{matter}(\tilde{\Psi},e^a_{\mu})=0$, where $\tilde{S}_{matter}$
is matter action of background fields ($L_c \tilde{\Psi}=S\tilde{\Psi}=0$).
Then the condition of topological symmetry breaking is the Einstein equation with 
matter,
\be
\frac{\delta e(R+\tilde{S}_{matter})}{\delta e^a_{\mu}}
=e\,e^a_{\mu}(R+\tilde{S}_{matter})+e({R^a}_{\mu}-{\tilde{T}^a}_{\mu})=0
\ee

and the torsion equation ,
\be
\frac{\delta e(R+\tilde{S}_{matter})}{\delta w^{ab}_{\mu}}= 
-6e \, e^{\mu}_{[a} e^{\nu}_l e^{\lambda}_{b]} (D_{\nu} e^l_{\lambda})
-{\tilde{S}^{\mu}}_{ab}=0
\ee
,where ${\tilde{T}^a}_{\mu}=-\frac{\delta \tilde{S}_{matter}}{\delta{e_a^{\mu}}}$ is a energy momentum tensor and ${\tilde{S}^{\mu}}_{ab}=
\frac{\delta \tilde{S}_{matter}}{\delta w^{ab}_{\mu}} $ is a spin density. 
To satisfy these (57) and (58), we have to change the gauge conditions of
the torsion free condition, eq.(15), and instead of eq.(14) 
$R_{ab}=0$ , we need 
\be
{R^a}_{\mu}-{\tilde{T}^a}_{\mu}=0
\ee

Since we have fixed $R+\tilde{S}_{matter}(\tilde{\Psi},e^a_{\mu})=0$, then
the following equation is necessary for symmetry breaking, 
\be
\tilde{S}_{matter}={\tilde{T}^a}_{\mu}e^{\mu}_a= tr({\tilde{T}^a}_{\mu}).
\ee

For example, Dirac field $\tilde{S}_{matter}=\bar{\tilde{\Psi}}
{\gamma}^{\mu} {\nabla}_{\mu} \tilde{\Psi}$
satisfy this condition. 
Then the topological symmetry breaking occurs depending upon matter fields. 
We may construct the theory that break topological symmetry by 
dynamics of matter fields. \\

 This study has constructed topological-like field theory that has 
broken phase when the solution of the Einstein equations exist.
This symmetry breaking is caused by Gribov zero modes, in other words by 
zero eigen values of Jacobiaan matrix, that appear as solutions of the 
Einstein equations. 
And we found that if one can take gauge fermion  cohomological
 of reduced BRS operator $\hat{S}$, then we can induce new BRS operator
 by the reduced symmetry.
In our case, we got the ${L_c}'$ as a new BRS operator and only symmetry of 
diffeomorphism is left. 
To take away the divergence from zero modes, we gave one method of regularization. 
Hence we could extend the topological gravity which was fixed on 
$R+\alpha=0 \; (\alpha >0)$
for conformal symmetry to the topological-like gravity on $R=0$. 
It has nontrivial broken phase on some manifolds, and especially when we chose the 
theory as eq.(54) and (55) then the theory describe
 semiclassical Einstein gravity. Of course, this theory dose not described 
the real gravity perfectly, as it is. But the property that breaking 
topological symmetry depend on back ground manifolds 
encourages us to apply our theory  to other theories.   
For example, we will have to examine the same methods to Weyl gravity. 
In our theory, only the conformal symmetry was needed to break the BRS symmetry. 
So, we might carry out the same methods easily in the Weyl gravity. 
We might construct quantum gravitational thory that 
have phase of more real seniclassical Einstein gravity. 
While the same phenomena will be realized in other theory as well, 
for example in topological Yang-Mills theory. \\
In this formalism, the gauge condition directly reflect broken 
phase physics.
There could be some criticisms. First, the way of breaking has
many ambiguities of selecting breaking terms. 
Second, it is unnatural that 
we have to adopt $\hat{S}$ cohomological gauge fermion for inducing 
the 2nd BRS operator.  But it may be interpreted as follows. 
If many BRS theory is found in real world, as we had seen in section 3, 
it's gauge condition will be restricted as cohomology of reduced BRS. 
It implies that internal space, i.e.gauge space, 
have some mechanism or kinematics.
It may be that as a result of it, gauge condition is non free from physics.
Further, ghost fields become phisical fields after symmetry breaking. 
We have to adjust and interpretate these new phisical fields to real world.\\ 

{\bf \large Acknowledge.}\\
 I am grateful to Professor Isikawa and Assistant Professor Suzuki for 
helpful suggestions and observations and a critical reading of the manuscript.


\end{document}